\documentclass[12pt,preprint]{aastex}
\shorttitle{SMA and Spitzer Observations of L1448 IRS2E}%
\shortauthors{Chen et al.}%

\begin{document}

\title{L1448 IRS2E: A CANDIDATE FIRST HYDROSTATIC CORE}

\author{Xuepeng~Chen\altaffilmark{1}, H\'{e}ctor~G.~Arce\altaffilmark{1},
Qizhou~Zhang\altaffilmark{2}, Tyler~L.~Bourke\altaffilmark{2},
Ralf~Launhardt\altaffilmark{3}, Markus~Schmalzl\altaffilmark{3},
and Thomas~Henning\altaffilmark{3}}

\affil{$^1$Department of Astronomy, Yale University, Box
208101, New Haven, CT 06520-8101, USA}
\affil{$^2$Harvard-Smithsonian Center for Astrophysics, 60
Garden Street, Cambridge, MA 02138, USA}
\affil{$^3$Max Planck Institute for Astronomy,
K\"{o}nigstuhl 17, D-69117 Heidelberg, Germany}

\begin{abstract}

Intermediate between the prestellar and Class\,0 protostellar
phases, the first core is a quasi-equilibrium hydrostatic object
with a short lifetime and an extremely low luminosity. Recent MHD
simulations suggest that the first core can even drive a molecular
outflow before the formation of the second core (i.e., protostar).
Using the Submillimeter Array and the {\it Spitzer Space
Telescope}, we present high angular resolution observations
towards the embedded dense core IRS2E in L1448. We find that
source L1448 IRS2E is not visible in the sensitive $Spitzer$
infrared images (at wavelengths from 3.6 to 70\,$\mu$m), and has
weak (sub-)\,millimeter dust continuum emission. Consequently,
this source has an extremely low bolometric luminosity
($<$\,0.1\,$L_\odot$). Infrared and (sub-)\,millimeter
observations clearly show an outflow emanating from this source;
L1448 IRS2E represents thus far the lowest luminosity source known
to be driving a molecular outflow. Comparisons with prestellar
cores and Class\,0 protostars suggest that L1448 IRS2E is more
evolved than prestellar cores but less evolved than Class\,0
protostars, i.e., at a stage intermediate between prestellar cores
and Class\,0 protostars. All these results are consistent with the
theoretical predictions of the radiative/magneto hydrodynamical
simulations, making L1448 IRS2E the most promising candidate of
the first hydrostatic core revealed so far.

\end{abstract}


\section{INTRODUCTION}

Stars form by the gravitational collapse of dense cores in
molecular clouds. A comprehensive understanding of the formation
and evolution of dense cores is thus a necessary prerequisite to
the understanding of the origin of stellar masses, multiple
systems, and outflows. Over the past decade, observational studies
of (low-mass) dense cores have made significant progress (see,
e.g., Reipurth et al. 2007 for recent reviews). Representing the
earliest phase of star formation, both prestellar and protostellar
cores have been extensively observed and studied using large
(sub)\,millimeter telescopes (e.g., JCMT and IRAM-30m) and
infrared telescopes (e.g., {\it Spitzer Space Telescope}). In
practice, however, it is still difficult to distinguish the two
types of cores because of the lack of readily observable
differences between them. This is illustrated by the fact that
several ``prestellar" cores, like L1014, were found to harbor very
low-luminosity protostars in sensitive $Spitzer$ observations (see
Young et al. 2004). Consequently, despite all of the observational
advances in the past decade, we still do not have a good
understanding of the evolutionary process that turns a prestellar
core into a protostar.

On the theoretical side, the collapse and evolution from
prestellar cores to Class\,0 protostars have been long studied
since the pioneering work of Larson (1969). Theoretical
calculations and simulations in fact predict two successive
collapse phases, before and after the dissociation of molecular
hydrogen, resulting in two different hydrostatic objects (see,
e.g., Larson 1969; Masunaga et al. 1998, 2000; Andr\'{e} et al.
2008). The collapsing prestellar core is initially optically thin
to the thermal emission from dust grains, and the compressional
heating rate by the collapse is much smaller than the cooling rate
by the thermal radiation. The collapse is therefore isothermal at
the very beginning. This condition is broken when the
compressional heating rate surpasses the radiative cooling rate,
and the central temperature increases gradually above 10\,K. The
collapse is then decelerated and forms a shock at the surface of a
quasi-adiabatic hydrostatic object, the so-called ``first
hydrostatic core" or ``first core", which consists mainly of
hydrogen molecules. The inward motion at this phase is called the
``first collapse". When the central temperature reaches about
2000\,K, hydrogen molecules begin to dissociate into atoms, which
acts as an efficient coolant of the gas. When released
gravitational energy is consumed by the dissociation, the gas
pressure cannot increase rapidly enough to support the first core
against its self-gravity, the ``second collapse" begins. After the
dissociation is completed, the ``second core", a truly hydrostatic
protostellar object, forms in the center. Most, if not all,
Class\,0 protostars observed so far actually belong to the
population of the ``second cores" (Ph. Andr\'{e}, private
communication).

Intermediate between the prestellar and Class\,0 protostellar
phases, the first core is a transient object accreting from the
surrounding dense envelope; the lifetime of the first core is
calculated to be only 10$^3$ to 10$^4$ years (Boss \& Yorke 1995;
Masunaga et al. 1998; Machida et al. 2008). Based on radiative
hydrodynamical (RHD) simulations, Boss \& Yorke (1995) and
Masunaga et al. (1998) modelled the spectral energy distribution
(SED) of the first core, and found that it should have an
extremely low bolometric luminosity ($<$\,0.1\,$L_\odot$), and
have no detectable infrared emission at wavelengths shorter than
$\sim$\,30\,$\mu$m with current telescopes. Furthermore, recent
magneto-hydrodynamical (MHD) simulations have found that the first
core can even drive a molecular outflow before the formation of
the second core (i.e., protostar) (Tomisaka 2002; Banerjee \&
Pudritz 2006; Machida et al. 2008). Therefore, the observational
detection of the first core would not only confirm the predictions
of RHD models but also set strong constraints on MHD models of
protostellar outflows. Unfortunately, due to its short lifetime
and extremely low luminosity, no first core has been
observationally found as yet.

In this paper, we present Submillimeter Array\footnote{The
Submillimeter Array is a joint project between the Smithsonian
Astrophysical Observatory and the Academia Sinica Institute of
Astronomy and Astrophysics and is funded by the Smithsonian
Institution and the Academia Sinica.} (SMA; Ho et al. 2004) and
{\it Spitzer Space Telescope} ($Spitzer$) observations towards an
embedded dense core in the L1448 region
($d$\,=\,240\,$\pm$\,20\,pc; Hirota et al. 2008). As a bridge
between the isolated star-forming cores and the large-scale
clusters, L1448 is an excellent region for studying star formation
on the intermediate scale and has been observed extensively in the
past two decades (see, e.g., Bally et al. 2008 and reference
therein). L1448~IRS2, in the western part of the L1448 filament,
was classified as a Class\,0 protostar by O'Linger et al. (1999).
Located $\sim$\,50$''$ to the east of IRS2, another dense core was
revealed in the SCUBA submm images in O'Linger et al. (1999), and
was formally cataloged as SCUBA core No.\,31 in Hatchell et al.
(2005) and SMM\,J032543+30450 in Kirk et al. (2006). This core was
found to have a mean kinetic gas temperature of $T_{\rm
kin}\approx 11$\,K, and the observed width of NH$_{3}$\,(1,\,1) is
$\sim$\,0.16\,km\,s$^{-1}$ (Rosolowsky et al. 2008). We refer to
this dense core as L1448 IRS2E in this work.

\section{OBSERVATIONS AND DATA REDUCTION}

\subsection{SMA Observations}

L1448 IRS2E was observed with the SMA on 2009 December 25th in the
compact configuration (seven antennas were used in the array). The
digital correlator was set up to cover the frequency ranges
216.9$-$220.9\,GHz and 228.8$-$232.8\,GHz in the lower and upper
sidebands, respectively. The three isotopic CO\,(2--1) lines and
several other lines, e.g., N$_{2}$D$^{+}$\,(3--2) and SiO\,(5--4),
were observed simultaneously in this setup. The 1.3\,mm dust
continuum emission was also recorded with a total bandwidth of
$\sim$\,7.5\,GHz ($\sim$\,3.8\,GHz USB and $\sim$\,3.7\,GHz LSB).
System temperatures ranged from 100 to 150\,K (depending on
elevation), with a typical value of $\sim$\,120\,K. Quasar 3c273
was used for bandpass calibration, and quasars 3c84 and 0359+509
for gain calibration. 3c273 was also used for absolute flux
calibration, from which we estimate a flux accuracy of
$\sim$\,20\%, by comparison of the final quasar fluxes with the
SMA calibration database. The data were calibrated using the IDL
MIR package and imaged using the Miriad toolbox (Sault et al.
1995). The SMA synthesized beam size and theoretical noise levels
at 1.3\,mm dust continuum and in the $^{12}$CO\,(2--1) line, with
robust {\it uv} weighting 1.0, are 3.9$''$\,$\times$\,2.6$''$,
0.48\,mJy\,beam$^{-1}$, and $\sim$\,53\,mJy\,beam$^{-1}$ (channel
width $\sim$\,1.0\,km\,s$^{-1}$), respectively.

\subsection{$Spitzer$ Observations}

The infrared data for L1448 were obtained from the $Spitzer$
Science Center (SSC) archive. The L1448 dark cloud, which is part
of the Perseus molecular cloud, was observed in 2004 as part of
the $Spitzer$ Legacy Program ``From Molecular Cores to Planet
Forming Disks" (c2d; Evans et al. 2003), by both the Infrared
Array Camera (IRAC) and the Multiband Imaging Photometer for
$Spitzer$ (MIPS). The infrared data were reduced by the c2d team
and are publicly available from the SSC science archive. The IRAC
and MIPS results of Perseus have been published by J{\o}rgensen et
al. (2006) and Rebull et al. (2007), respectively. The details of
the observations and data reductions can be found in Dunham et al.
(2008).

\section{RESULTS}

\subsection{Infrared, Submillimeter, and Millimeter Continuum Emission}

Figure~1 shows the SCUBA 850\,$\mu$m dust continuum contours of
the L1448 complex (from Kirk et al. 2006; publicly available on
the COMPLETE web site\footnote{http://cfa.harvard.edu/COMPLETE}),
plotted on the $Spitzer$ images. The SCUBA contours show the
arc-shaped filamentary structure of L1448, in which the three
well-known Class\,0 protostars (i.e., L1448C, IRS3, and IRS2) are
labelled. Located $\sim$\,50$''$ to the east of source IRS2, the
dense core, referred to as IRS2E, is seen in the SCUBA image. This
core is also spatially coincident with a molecular cloud core
revealed by IRAM-30m N$_{2}$H$^{+}$\,(1--0) observations (Chen et
al. 2010, in preparation), and has a line-of-sight velocity (LSR
velocity $\sim$\,4.1\,km\,s$^{-1}$; Rosolowsky et al. 2008)
similar to the other Class\,0 protostars, indicating it is
physically associated with the L1448 filament. However, no compact
infrared emission was detected from L1448 IRS2E in any of the
$Spitzer$ bands (from 3.6 to 70\,$\mu$m; see Figs.\,1a-1d),
suggesting that this source is extremely cold. We note that the
$Spitzer$ data of L1448 were obtained with the c2d legacy program,
a large infrared survey toward five nearby large molecular clouds
(Chamaeleon\,II, Lupus, Ophiuchus, Perseus, and Serpens). In the
c2d observations, all the known Class\,0 protostars in these five
clouds were detected (Evans et al. 2009 and references therein),
which proves that these images have enough sensitivity to
detect the youngest protostars. This indicates that the non-detection 
of L1448 IRS2E in the $Spitzer$ images is not due to insufficient 
imaging sensitivity but is observationally significant.

In the SMA 1.3\,mm dust continuum images, a weak continuum source
($\sim$\,6\,$\sigma$ level; R.A.\,=\,03:25:25.66,
Dec.\,=\,30:44:56.7, J2000) is found within the IRS2E core (see
Figs.\,1b-1d). This source is evident in images made from
visibility data of either sideband, as well as in both halves of
the track, suggesting that it is not an artifact in the data.
Interestingly, the IRAC images (at all four bands) show a diffuse
jet-like feature to the south of IRS2E, with this SMA dust
continuum source located at the apex (see Figs.\,1a-1b). From
Gaussian fitting in the {\it cleaned-restored} images, we derive a
flux density of 6\,$\pm$\,2\,mJy for this dust continuum source.
Assuming that the 1.3\,mm dust continuum emission is optically
thin, the total gas mass ($M_{\rm gas}$) of IRS2E is calculated
with the same method described in Launhardt \& Henning (1997). In
the calculations, we adopt a dust opacity of $\kappa_{\rm m} =
0.5\,{\rm cm}^2\,{\rm g}^{-1}$, which is a typical value for cold
and dense cores with an average number density of $n(\rm H) =
10^5\,$cm$^{-3}$ (Ossenkopf \& Henning 1994), and a dust
temperature of $\sim$\,11\,K, which is similar to the kinetic gas
temperature of L1448 IRS2E. The total gas mass of this source,
estimated from the SMA dust continuum observations, is
$\sim$\,0.04\,$\pm$\,0.01\,$M_\odot$.

To further estimate the submm and mm fluxes, in Figure~2 we show
the SCUBA 450\,$\mu$m, 850\,$\mu$m, and Bolocam 1.1\,mm images of
L1448 IRS2, taken from the JCMT science archive\footnote{Because
the 850\,$\mu$m fluxes in Kirk et al. (2006; 2007) are somewhat
uncertain (due to issues with the calibration), we prefer using
the original SCUBA images of L1448 IRS2 (from the JCMT science
archive) to estimate the submm fluxes of L1448 IRS2E.} and Enoch
et al. (2006), respectively. As shown in Fig.\,2, the submm
continuum emission from source IRS2E shows a roughly
centrally-peaked condensation separated from the IRS2 core, while
the 1.1\,mm continuum emission from IRS2E is weak and elongated,
and no core was identified at this position by Enoch et al.
(2006). The fluxes within one beam around the SMA continuum source
are estimated to be $\sim$\,1200\,mJy at 450\,$\mu$m,
$\sim$\,400\,mJy at 850\,$\mu$m, and $\sim$\,300\,mJy at 1.1\,mm,
respectively (see Fig.\,2 and Table~1). However, it must be noted
that the peak positions of the IRS2E core in the SCUBA
450\,$\mu$m, 850\,$\mu$m, and SMA 1.3\,mm images are different
from each other ($\sim$\,10$''$ offset). It appears that the
single-dish data reveal the entire (extended) low-mass dense core,
while the SMA data shows a faint compact source which is embedded
in this core, but not at its center. Deeper (sub-)\,mm dust
continuum observations are needed to investigate the position and
density structure of this core.

\subsection{CO\,(2--1) Emission}

Figure~3 shows the velocity channel maps of the SMA
$^{12}$CO\,(2--1) emission of L1448 IRS2E. For comparison, the SMA
$^{12}$CO\,(2--1) emission from L1448 IRS2 is also plotted here
(the SMA data of IRS2 were taken from Arce et al. 2010, in
preparation). For IRS2E, CO emission is detected to the south of
the SMA continuum peak at velocities from $V_{\rm
LSR}$\,=\,$\sim$\,7 to $\sim$\,30\,km\,s$^{-1}$, with the cloud
systemic velocity being $\sim$\,4.1\,km\,s$^{-1}$. This redshifted
CO emission exhibits an elongated and narrow structure. No
blueshifted CO emission was detected around IRS2E in the SMA
observations, even though the SMA primary beam of $\sim$\,55$''$
(at 230\,GHz) covers the area where we would expect to see the
blueshifted lobe if it were there. Hatchell et al. (2007) observed
the L1448 region in the CO\,(3--2) line using the JCMT, and
similar to our findings, they detected no blueshifted emission
around L1448 IRS2E.

As shown in Fig.\,3, several elongated CO lobes are seen around
IRS2E in the velocity channels from $\sim$\,7 to $\sim$\,13\,km\,s$^{-1}$,
which come from the redshifted outflow driven by L1448 IRS2. The
Class\,0 protostar IRS2 was first proposed to be a binary system by
Wolf-Chase et al. (2000) based on the NRAO-12\,m CO\,(1--0)
observations that mapped two distinct outflows from IRS2. 
Volgenau et al. (2002) claimed the detection of a binary system in 
IRS2 using BIMA, but no results (e.g., spatial separation and image) 
are reported as yet. Other studies (e.g., Tobin et al. 2007) have 
referenced Volgenau et al. (2002) and Volgenau  (2004) as stating 
that IRS2 is part of a binary system with a  companion with about 
10\,$''$ to the northwest. We would have expected to detect this 
presumed companion in our SMA images, which have a synthesized 
beam of $\sim$\,3$''$ (see Figure 4),  but we do not see evidence in 
our SMA dust continuum maps. Moreover, we do not find any evidence 
of this presumed companion in the IRAC and MIPS images, nor it is 
detected in the 450\,$\mu$m SCUBA map (with a $\sim$\,8$''$ beam, 
see Fig.\,2). We therefore consider IRS2 as a single source here. In 
fact, the structures of the outflow from IRS2 can be explained without 
the need to invoke the unseen binary companion (see Figures 3 and 4). 
Most of the structures seen at  redshifted velocities appear to 
delineate the walls of the 50 deg-wide outflow cavities, while the 
elongated redshifted emission along the outflow axis most probably 
traces the dense collimated part of the outflow, as seen in other 
sources (e.g., Santiago-Garc\'{i}a et al. 2009) and predicted by 
so-called `unified' outflow model (e.g., Shang  et al. 2007) (see 
Fig.\,4). More details about the IRS2 outflow will be
presented in another paper (Arce et al., in preparation).

Figure~4 shows the velocity-integrated intensity map of the SMA
$^{12}$CO\,(2--1) emission of L1448 IRS2E and IRS2. To the south
of IRS2E, the collimated redshifted CO lobe ($\sim$\,40$''$ or
9600\,AU in length) is spatially coincident with the infrared jet
detected in the $Spitzer$ IRAC images and the CO\,(3--2) red
emission detected at the JCMT. The orientation and morphology of
this CO lobe suggest that it is neither part of the cavity wall of
the IRS2 extended outflow nor part of the molecular jet from IRS2. 
These results indicate that L1448 IRS2E, a cold core with no 
detectable infrared emission, is driving a molecular outflow. 
Nevertheless, further observations, e.g., short-spacing data, are 
needed to recover the missing flux of extended structure, and to 
improve the quality of the outflow maps.

Assuming that the $^{12}$CO\,(2--1) line emission is optically
thin, the outflow mass of L1448 IRS2E is derived with the standard
manner (e.g., Cabrit \& Bertout 1990). In the calculations, we
assume LTE conditions and an excitation temperature of 20\,K (the
values in the range of 10$-$50\,K modify the calculations by less
than a factor of 2). The derived outflow mass of IRS2E is about
2\,$\times$\,10$^{-3}$\,$M_\odot$. For other properties relying on
a knowledge of the outflow velocity (i.e., age $\tau_{\rm flow}$,
momentum $P$, energy $E$, force $F_{\rm m}$, and mechanical
luminosity $L_{\rm m}$), we adopt a value of 25\,km\,s$^{-1}$,
where we assume that the outflowing gas is moving at the maximum
observed velocity. We obtain $\tau_{\rm flow}$\,$\sim$\,1800\,yr
(assuming a lobe size of 9600\,AU),
$P$\,$\sim$\,0.05\,$M_\odot$\,km\,s$^{-1}$,
$E$\,$\sim$\,1.2\,$\times$\,10$^{43}$\,ergs, $F_{\rm
m}$\,$\sim$\,2.5\,$\times$\,10$^{-5}$$M_\odot$\,km\,s$^{-1}$\,yr$^{-1}$,
and $L_{\rm m}$\,$\sim$\,0.05\,$L_\odot$, without correcting for
the unknown inclination of the outflow with respect to the plane
of the sky. The outflow mass-loss rate $\dot{M}$$_{\rm out}$,
estimated directly from the mass and age $\tau_{\rm flow}$, is
$\sim$\,1.0\,$\times$\,10$^{-6}$\,$M_\odot$\,yr$^{-1}$. We note
that all these outflow parameters refer only to the compact
outflows detected in the SMA maps and thus represent lower limits.

\section{DISCUSSION}

\subsection{Spectral Energy Distribution}

Table~1 lists the (sub-)\,mm fluxes of L1448 IRS2E, estimated from
the SCUBA, Bolocam, and SMA images. Since there is no local
emission peak at the position of the SMA compact source in the
SCUBA/Bolocam images (see Fig.\,2), the estimated fluxes per beam
around IRS2E in these images represent conservative upper limits
to the fluxes from the embedded source. The 3\,$\sigma$ upper
limits in the $Spitzer$ images are also listed in Table~1. Based
on these data points, we constructed the spectral energy
distribution (SED) of IRS2E (plot not shown here). To
estimate the luminosity of IRS2E, we first interpolated and then
integrated the SED (all the upper limits were used), always
assuming spherical symmetry. Interpolation between the flux
densities was done by a $\chi$$^2$ single-temperature grey-body
fit to all points at $\lambda$\,$\geq$\,70\,$\mu$m, using the same
method as described in Chen et al. (2008). A simple logarithmic
interpolation was performed between all points at
$\lambda$\,$\leq$\,70\,$\mu$m. The estimated bolometric luminosity
of L1448 IRS2E is less than 0.1\,$L_\odot$.

Although only an upper limit to the bolometric luminosity could be
derived, we can still use it to further constrain the evolutionary
stage of L1448 IRS2E. If we assume a steady mass-accretion rate
given by $\dot{M}$\,=\,0.975$c_{\rm s}^3$/$G$ (Shu 1977), where
$c_{\rm s}$ is the effective sound speed, for a gas temperature of
10\,K the accretion rate is
$\sim$\,2\,$\times$\,10$^{-6}$\,$M_\odot$\,yr$^{-1}$. The
accretion luminosity is calculated as $L_{\rm
acc}$\,=\,$GM_*\dot{M}$/$R_*$, where $M_*$ is the stellar mass and
$R_*$ is the stellar radius. The bolometric luminosity being
$<$\,0.1\,$L_\odot$ implies a protostellar mass of
$<$\,0.01\,$M_\odot$, assuming a radius of 2\,$R_\odot$. The age
of a $<$\,0.01\,$M_\odot$ `protostar' under the assumption of a
constant mass-accretion rate of
2\,$\times$\,10$^{-6}$\,$M_\odot$\,yr$^{-1}$ is then calculated to
be $<$\,5000\,yr, which is consistent with the outflow age
estimated above ($\geq$\,1800\,yr).

The estimated low luminosity and age suggest that L1448 IRS2E is a
very young object, in which star formation has just started.
Nevertheless, it must be noted that uncertainties remain in our
estimates due to the limited observations available. More
information, such as {\it Herschel Space Observatory} imaging at
75$-$300\,$\mu$m, is needed to constrain the SED of L1448 IRS2E in
order to address more precisely its evolutionary status.

\subsection{Comparisons to Prestellar, Class\,0, and VeLLO Objects}

\noindent{\bf Comparison to Prestellar Cores:}
Prestellar cores are dense ($n$$_{\rm
H}$\,$\sim$\,10$^{4}$--10$^{6}$\,cm$^{-3}$) cores which are
self-gravitating and evolve toward higher degrees of central
condensation, but no central hydrostatic protostellar object
exists yet within the core (Andr\'{e} et al. 2000; 2008). Although
the properties of L1448 IRS2E are still poorly known, its observed
narrow width of the NH$_{3}$ line ($\sim$\,0.16\,km\,s$^{-1}$;
Rosolowsky et al. 2008), as well as the fact that no point-like
source is detected in the $Spitzer$ images, resemble the
properties of prestellar cores (see Andr\'{e} et al. 2008).
However, as suggested by the SMA CO\,(2--1) observations, L1448
IRS2E appears to drive a molecular outflow, which implies ongoing
accretion onto a central condensation and has never seen before in
prestellar cores. Furthermore, the estimated ratio of
$I$[C$^{18}$O(1$-$0)] (Hatchell et al. 2005) to
$I$[N$_2$H$^+$(1$-$0)] (Kirk et al. 2007) in the IRS2E core is
$\sim$\,0.26, similar to that of `evolved' prestellar cores, like
L1544 (see Tafalla 2005), which suggests that the IRS2E core is
chemically evolved and probably already passed the last stage of
the prestellar phase.

\noindent{\bf Comparison to Class\,0 Objects:}
Class\,0 objects are the youngest accreting protostars with an age
of a few~$\times$\,10$^{4}$\,yr. These objects are in an early
evolutionary stage, right after point mass formation, when most of
the mass of the system is still in the surrounding dense
core/envelope (Andr\'{e} et al. 2000). They represent the truly
hydrostatic protostellar objects (i.e., the second core) formed in
dense cores. So far at least 50 Class\,0 protostars have been
identified (Andr\'{e} et al. 2000; Froebrich 2005). Most of them
are detectable in the $Spitzer$ images (at least in the MIPS
bands), and are associated with strong submm and mm dust continuum
emission (in both single-dish and interferometric maps). Although
the collimated outflow from IRS2E possesses the typical properties
of an outflow from a Class\,0 protostar (see Arce et al. 2007), an
obvious difference between L1448 IRS2E and known Class\,0
protostars (e.g., L1448C, IRS3, and IRS2) is that IRS2E is not
visible in the sensitive $Spitzer$ images, has weak dust continuum
emission, and consequently has an extremely low bolometric
luminosity ($<$\,0.1\,$L_\odot$). The estimated age of L1448 IRS2E
(a~few\,$\times$\,10$^{3}$\,yr) is also much less than those of
the Class\,0 protostars, suggesting that IRS2E is younger
(less-evolved) than Class\,0 protostars.

Furthermore, we compare L1448 IRS2E to another source in the 
Perseus molecular cloud: SVS\,13B (see Chen et al. 2009 and 
references therein). Like L1448 IRS2E, SVS\,13B has no point-like 
infrared emission at wavelengths from 3.6 to 70\,$\mu$m in the 
$Spitzer$ images (also c2d data). However, it must be noted that 
SVS\,13B is located $\sim$\,15$''$ to the south of the bright Class\,I 
object SVS\,13A, and thus the detection limits in the $Spitzer$ 
images around SVS\,13B are about three times worse than those 
in the L1448 images (because the imaging backgrounds around 
SVS\,13B were raised by the bright source SVS\,13A). Interestingly, 
SVS\,13B is also driving a collimated outflow seen in the high angular 
resolution SiO and CO images (Bachiller et al. 1998; 2000). In 
contrast to L1448 IRS2E, SVS\,13B has much stronger dust continuum 
emission at submm and mm wavelengths, and correspondingly has 
much higher gas mass ($>$\,1\,$M_\odot$)  and bolometric luminosity 
($>$\,1\,$L_\odot$). In addition, the kinematic properties of SVS\,13B, 
e.g., fast rotation and subsonic turbulence (see Chen et al. 2009), are 
similar to those of Class\,0 protostars (e.g., Chen et al. 2007). Therefore, 
SVS\,13B is very likely more evolved than L1448 IRS2E and has already
formed an extremely young Class\,0 protostar.

\noindent{\bf Comparison to Known VeLLOs:}
The extremely low luminosity of L1448 IRS2E is similar to what is
seen in the so-called very low luminosity objects (VeLLOs), an
interesting subset of embedded, low-luminosity protostars (see
Dunham et al. 2008 and references therein). However, non-detection
at both 24 and 70\,$\mu$m bands distinguishes L1448 IRS2E from all
VeLLOs revealed thus far (Dunham et al. 2008). Direct
observations, together with radiative transfer modelling, have
shown that young (sub-)\,stellar objects have already formed in
these VeLLOs. In contrast, there is yet no clear evidence for the
presence of a protostar in L1448 IRS2E, even though the
sensitivities of the $Spitzer$ images of L1448 IRS2E are
comparable to those used to detect the known VeLLOs (see Dunham et
al. 2008 and references therein).

The evolutionary status and eventual final state of VeLLOs are still
unclear. Some of them, e.g., IRAM\,04191+1522 (see Dunham et al.
2006), represent typical Class\,0 low-mass protostars, while
others, e.g. L1014-IRS (see Bourke et al. 2005), could represent
precursors of sub-stellar objects (i.e., proto-brown dwarfs). In
the case of L1448 IRS2E, it is more likely that we are catching
the very first moments of low-mass star formation because L1448
IRS2E already has about 0.04\,$M_\odot$ of gas estimated from the
SMA dust continuum observations, and more gas in the outer
envelope/core can continue accreting onto it. If we assume a
steady accretion rate and a core-to-star efficiency of 15--30\%
(Evans et al. 2009), then it is very probable that a low-mass star
($\geq$\,0.1\,$M_\odot$) will eventually form in the L1448 IRS2E
core.

\subsection{A Candidate First Hydrostatic Core}

The observational detection of the first hydrostatic core is of
prime importance for understanding the early evolution of
star-forming dense cores and the origin of outflows. Encouraged by
these facts, searches for the first core have been undertaken over
the past decade. Based on the HCO$^+$/H$^{13}$CO$^+$ observations,
Onishi et al. (1999) suggested that L1521F could be a first core
candidate, but $Spitzer$ observations soon found that L1521F
harbors a low luminosity protostar (Bourke et al. 2006). More
recent studies suggest that the evolutionary stage of L1521F is
similar to or younger than the Class\,0 phase, and may be
consistent with the early second collapse phase (Shinnaga et al.
2009; Terebey et al. 2009). Another promising object was Cha-MMS1,
suggested by Belloche et al. (2006) from the measurement of the
deuterium fractionation. However, a mid-infrared source was
detected by $Spitzer$ MIPS observations, indicating a compact
hydrostatic object had already formed in Cha-MMS (see Belloche et
al. 2006 for more details).

Based on the SMA and $Spitzer$ observations, we find that source
L1448 IRS2E has the following characteristics: (1) it is not
visible in the sensitive $Spitzer$ infrared images (from 3.6 to
70\,$\mu$m); (2) has very weak (sub-)\,mm dust continuum emission,
and consequently has an extremely low bolometric luminosity
($<$\,0.1\,$L_\odot$); and (3) appears to drive a molecular
outflow. Comparisons with prestellar cores and Class\,0 protostars
suggest that L1448 IRS2E is more evolved than prestellar cores but
less evolved than Class\,0 protostars, i.e., at a stage
intermediate between prestellar cores and Class\,0 protostars.
These results are consistent with the theoretical predictions in
the RHD/MHD models for the first hydrostatic core (see
Section\,I)\footnote{In the MHD model of Machida et al. (2008),
the outflow driven by the first core has a slow speed of
$\sim$\,3\,km\,s$^{-1}$. However, in their models the first core 
only has a mass of 0.01\,$M_\odot$ at the end of the calculations. 
Since the first core will grow in mass by at least 1--2 orders of 
magnitude in the subsequent gas accretion phase, the relatively 
high outflow velocity of L1448 IRS2E ($\sim$\,25\,km\,s$^{-1}$; 
about 8 times larger than that in the model) could be explained by 
the relatively larger (gas) mass of the L1448 IRS2E core
($\sim$\,0.04\,$M_\odot$), i.e., a deeper gravitational potential
and a faster escape speed.}, making L1448 IRS2E the most promising
first hydrostatic core candidate thus far.

However, it must be noted that the nature of source L1448 IRS2E is
not definitive. More observations are needed to constrain its SED
and to refine its outflow maps. Detections of other objects like
L1448 IRS2E will be important for understanding the process of
dynamical collapse and the origin of outflows. Sensitive surveys
at wavelengths from far-infrared (e.g., $Herschel$) to (sub-)\,mm
continuum (e.g., SCUBA) are needed to search for more first core
candidates in nearby clouds. We also speculate that some of the
objects in the current sample of prestellar cores may already
harbor first cores, which drive molecular outflows hidden within
the extended cloud emission and are therefore not revealed in low
resolution single-dish observations. A systematic high-resolution
interferometric CO survey toward these cores is needed to search
for potential outflow activity.

\section{SUMMARY}

We present SMA and $Spitzer$ observations of the low-mass,
embedded dense core L1448~IRS2E. This core has no point-like
infrared emission in the $Spitzer$ images, and shows weak emission
in the SMA 1.3\,mm dust continuum map ($\sim$\,6\,mJy).
Consequently, it has an extremely low bolometric luminosity (less
than 0.1\,$L_\odot$). Interestingly, the SMA CO\,(2--1) images
suggest that L1448 IRS2E is driving a collimated CO outflow (up to
$\sim$\,25\,km\,s$^{-1}$), which is further supported by the
$Spitzer$ IRAC images with regards to the morphology of the
outflow. L1448 IRS2E represents so far the lowest luminosity
source with a detectable molecular outflow. A comparison with
prestellar cores and Class\,0 protostars suggests that L1448 IRS2E
is in an evolutionary stage between that of a prestellar core and
a Class\,0 protostar. Our results are consistent with the
predictions of the theoretical models for the first hydrostatic
core, making L1448 IRS2E thus far the most promising first 
hydrostatic core candidate. Further observations, such as {\it
Herschel Space Observatory} imaging at 75$-$300\,$\mu$m and
short-spacing CO observations, are needed to study its properties
and to address more precisely its evolutionary status. If the
properties of L1448 IRS2E are validated by further observations,
this would be the first confirmed detection of the first core
stage of star formation.

\acknowledgments
We thank the anonymous referee for many insightful comments and
suggestions. We acknowledge the SMA staff for technical support
during the observations and the $Spitzer$ Science Center for their
maintenance of the $Spitzer$ data.

\clearpage


\begin{deluxetable}{lccc}
\tabletypesize{\scriptsize} \tablecaption{\footnotesize Photometry
of L1448 IRS2E\label{photometry}} \tablewidth{0pt}%
\tablehead{\colhead{$\lambda$}&\colhead{$S_{\nu}$}&\colhead{Aperture}\\
\colhead{($\mu$m)}&\colhead{(mJy)}&\colhead{(arcsec)}}\startdata

3.6    & $<$\,0.021$^a$    & 1.2 \\
4.5    & $<$\,0.027$^a$    & 1.2 \\
5.8    & $<$\,0.16$^a$     & 1.2 \\
8.0    & $<$\,0.72$^a$     & 1.2 \\
24     & $<$\,18$^a$       & 2.5 \\
70     & $<$\,120$^a$      & 20  \\
160    & $<$\,2700$^a$     & 20  \\
450    & $<$\,1200$^b$     & 10  \\
850    & $<$\,400$^b$      & 14  \\
1100   & $<$\,300$^b$      & 30  \\
1300   & $>$\,6$^c$       & 3  \\

\enddata
\tablenotetext{a}{Detection upper limits (3\,$\sigma$ per
aperture) at the position of L1448 IRS2E in the $Spitzer$ images.}
\tablenotetext{b}{Estimations from the flux densities (mJy/beam)
within one beam around L1448 IRS2E in the JCMT/SCUBA and Bolocam
1.1\,mm images.} \tablenotetext{c}{~Lower limit detected in the
SMA images.}
\end{deluxetable}


\begin{figure*}[hlpt]
\begin{center}
\includegraphics[width=16cm,angle=0]{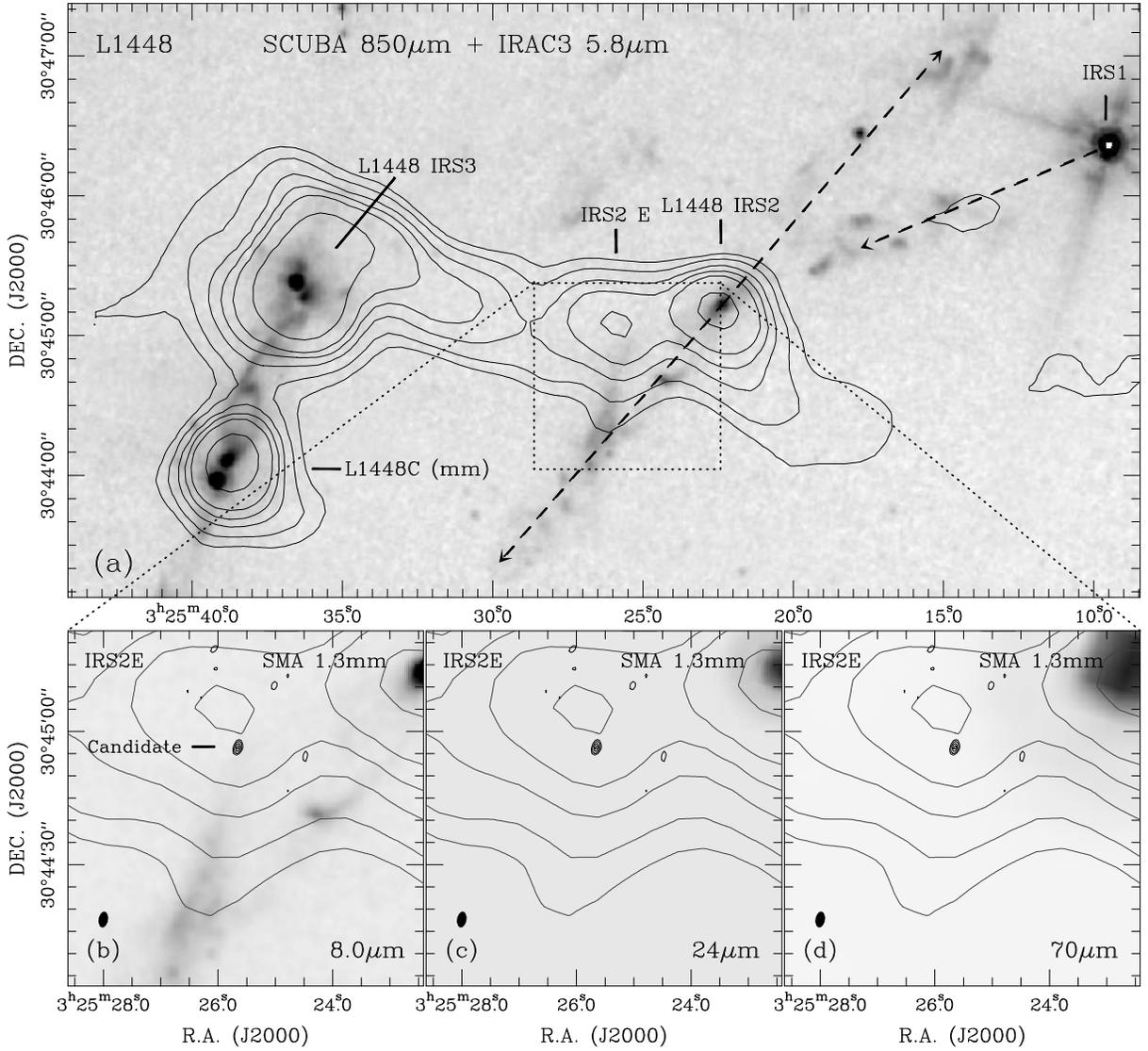}
\caption{(a) $Spitzer$ IRAC band\,3 (5.8\,$\mu$m) image of the
L1448 complex, overlaid with JCMT/SCUBA 850\,$\mu$m dust continuum
contours. The SCUBA 850\,$\mu$m dust continuum contours levels
correspond to 3, 5, 8, 12, 15, and 20\,$\sigma$, where 1\,$\sigma$
level is $\sim$\,40\,mJy\,beam$^{-1}$. Black dashed arrows show
the directions of the jets driven by IRS1 and IRS2, respectively
(see also Davis et al. 2008). (b) The SMA 1.3\,mm dust continuum
contours (black) of L1448 IRS2E, plotted on the $Spitzer$ IRAC
8.0\,$\mu$m image. The SMA contours start at $\sim$\,3\,$\sigma$
(1\,$\sigma$\,$\sim$\,0.85\,mJy\,beam$^{-1}$) with steps of
$\sim$\,1\,$\sigma$. The synthesized SMA beam is shown as a grey
oval in the bottom left corner. (c) The same as Figure~1b, but
plotted on the $Spitzer$ MIPS 24\,$\mu$m image. (d) The same as
Figure~1b, but for $Spitzer$ MIPS 70\,$\mu$m image.
\label{l1448irs2e_sma_spitzer}}
\end{center}
\end{figure*}


\begin{figure*}[hlpt]
\begin{center}
\includegraphics[width=16cm,angle=0]{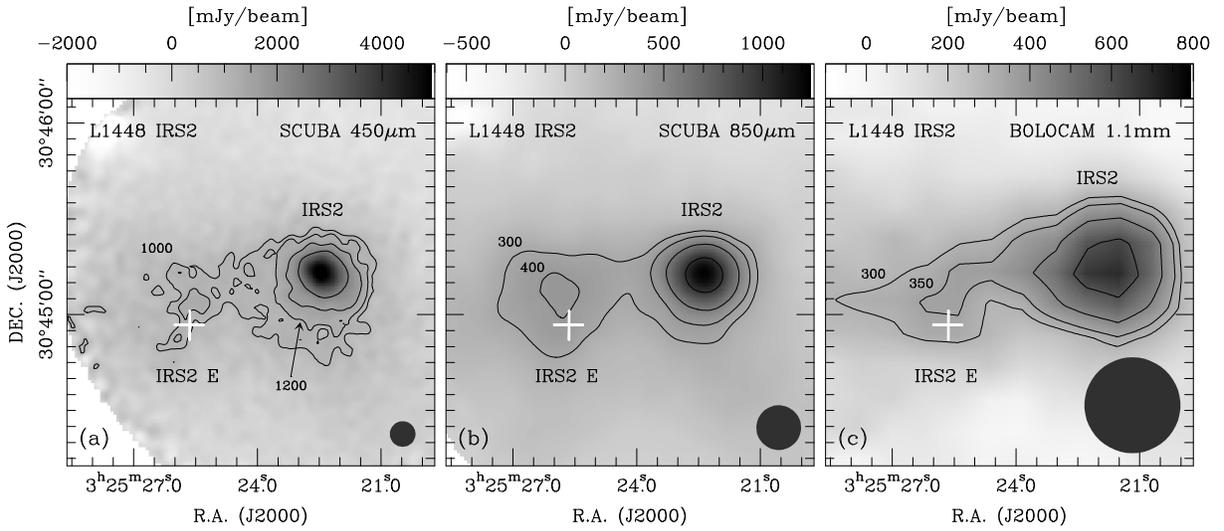}
\caption{(a) The SCUBA 450\,$\mu$m image of L1448 IRS2; Contour
levels correspond to 1000, 1200, 1600, and 2400\,mJy\,beam$^{-1}$.
(b) The SCUBA 850\,$\mu$m image; Contour levels 300, 400, 600,
1000\,mJy\,beam$^{-1}$. (c) The Bolocam 1.1\,mm image; Contour
levels 300, 350, 450, 600\,mJy\,beam$^{-1}$. In each image, the
filled ellipse indicates the FWHM of SCUBA and Bolocam, the
numbers indicate the value of the contour in mJy\,beam$^{-1}$,
while the cross marks the position of the SMA 1.3\,mm dust
continuum source associated with L1448 IRS2E.\label{l1448irs2e_sucba_bolocam}}
\end{center}
\end{figure*}

\begin{figure*}
\begin{center}
\includegraphics[width=16cm, angle=0]{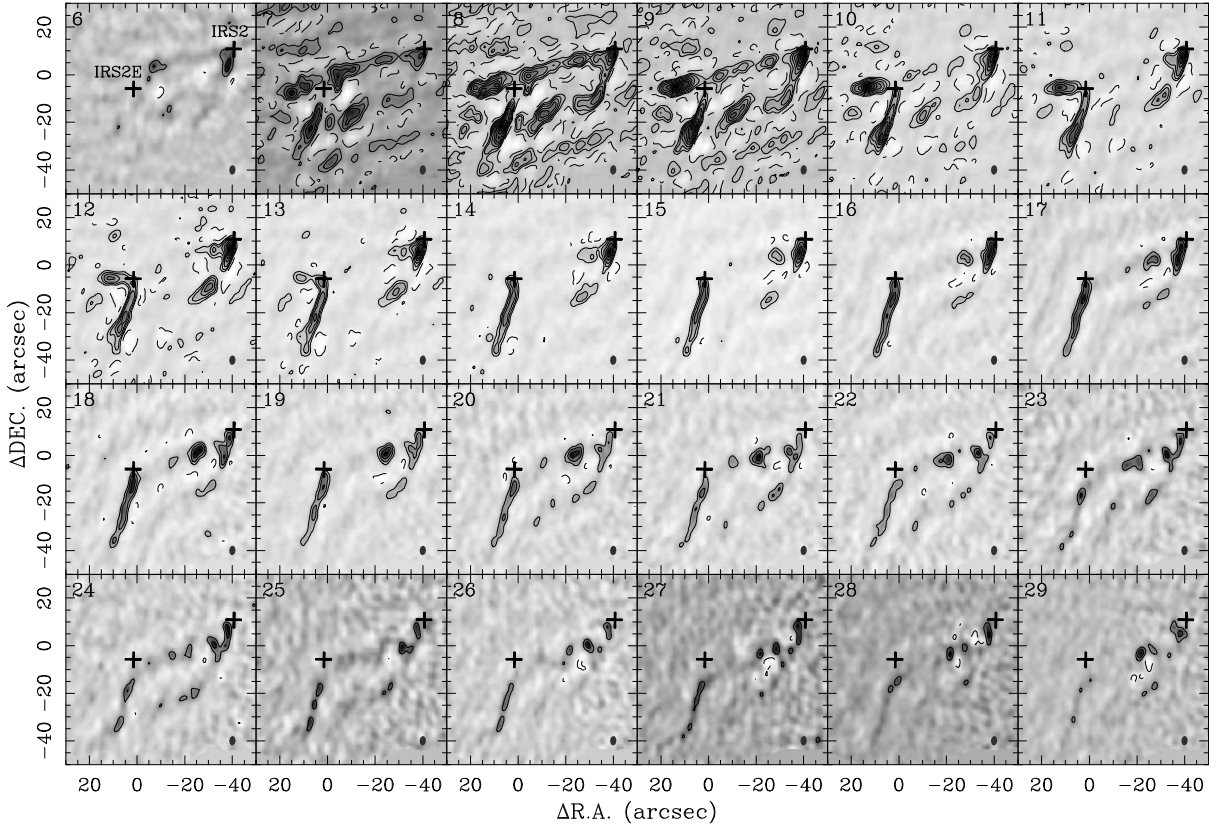}
\caption{Velocity channel maps of the SMA $^{12}$CO\,(2--1)
emission of L1448 IRS2E (phase center R.A.\,=\,03:25:25.52,
DEC\,=\,30:45:02.65, J2000). Contour levels correspond to $-$3, 3,
6, 10\,$\sigma$ then increase in steps of 5\,$\sigma$, where the
1\,$\sigma$ level is $\sim$\,0.1$-$0.15\,Jy\,beam$^{-1}$. In each
panel, the center velocity is written in the top left corner (in units of
km\,s$^{-1}$), the two crosses mark the positions of the SMA
1.3\,mm dust continuum sources of IRS2E and IRS2, and the filled
ellipse (lower right corner) indicates the synthesized beam of the
SMA.\label{l1448irs2e_outflow_channel}}
\end{center}
\end{figure*}

\begin{figure*}
\begin{center}
\includegraphics[height=12cm, angle=0]{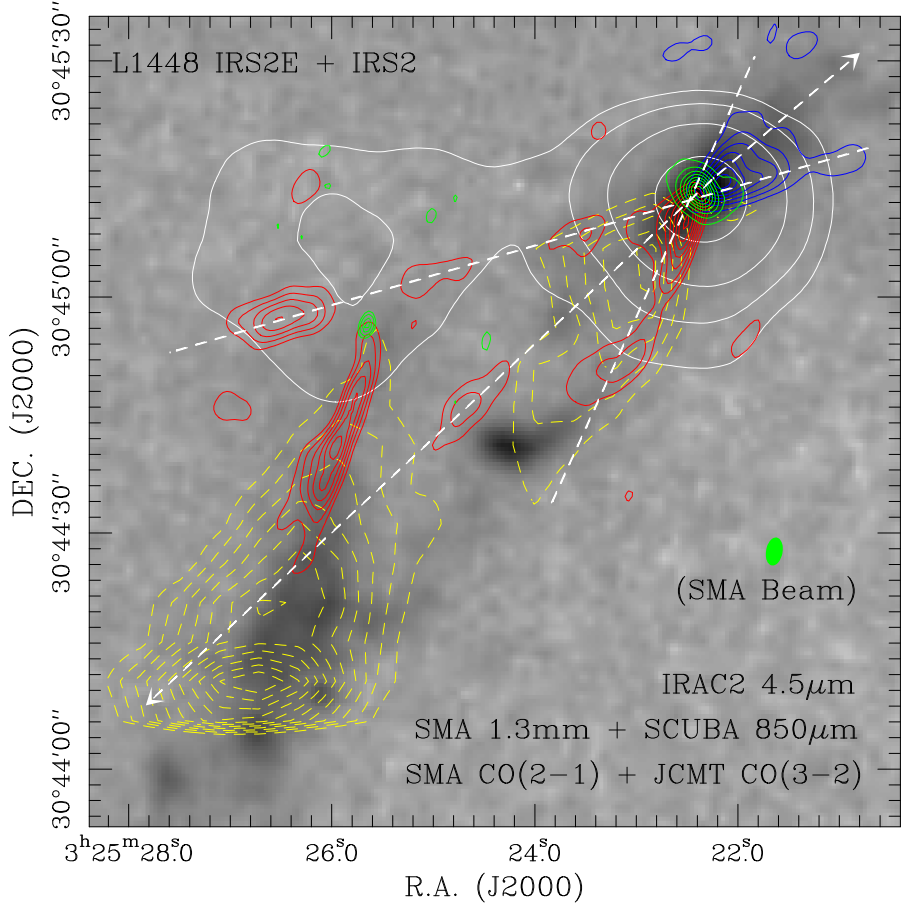}
\caption{Integrated intensity map of the SMA $^{12}$CO\,(2--1)
emission from L1448 IRS2E and IRS2, plotted on the $Spitzer$
4.5\,$\mu$m image. Red solid contours represent the SMA CO\,(2--1)
emission integrated over the velocity range
7\,km\,s$^{-1}$\,$<$\,$V_{\rm LSR}$\,$<$\,30\,km\,s$^{-1}$, which
is redshifted with respect to the cloud systemic velocity
($\sim$\,4.1\,km\,s$^{-1}$), while yellow dashed contours
represent the JCMT CO\,(3--2) emission integrated over the same
velocity range (from Hatchell et al. 2007).  Blue solid contours
represent the SMA CO\,(2--1) blueshifted emission from IRS2
integrated over the velocity range
$-$15\,km\,s$^{-1}$\,$<$\,$V_{\rm LSR}$\,$<$\,2\,km\,s$^{-1}$. The
redshifted SMA (JCMT) CO contours start at
$\sim$\,5.0\,Jy\,beam$^{-1}$\,km\,s$^{-1}$
($\sim$\,25\,K\,km\,s$^{-1}$), and increase in step of
4.0\,Jy\,beam$^{-1}$\,km\,s$^{-1}$ ($\sim$\,7\,K\,km\,s$^{-1}$),
while the blueshifted SMA CO contours start at
$\sim$\,2.0\,Jy\,beam$^{-1}$\,km\,s$^{-1}$, and increase in step
of 4.0\,Jy\,beam$^{-1}$\,km\,s$^{-1}$. The green and white
contours represent the SMA 1.3\,mm and the SCUBA 850\,$\mu$m dust
continuum emission, respectively. The synthesized SMA beam is shown 
as a green oval in the bottom right corner. For IRS2E, the SMA 1.3\,mm
contours are same as those in Figure\,1. For IRS2, the SMA 1.3\,mm
contours start at 0.01\,Jy\,beam$^{-1}$ and increase in step of
0.01\,Jy\,beam$^{-1}$ ($\sim$\,5\,$\sigma$). Dashed arrows show
the direction of the IRS2 outflow (central jet), and dashed lines
show the positions of the outflow cavity walls.\label{l1448irs2e_outflow}}
\end{center}
\end{figure*}

\end{document}